\documentclass[lettersize,journal]{IEEEtran}
\usepackage{amsmath,amsfonts}
\usepackage{algorithmic}
\usepackage{algorithm}
\usepackage{array}
\usepackage[caption=false,font=normalsize,labelfont=sf,textfont=sf]{subfig}
\usepackage{textcomp}
\usepackage{stfloats}
\usepackage{url}
\usepackage{verbatim}
\usepackage{graphicx}
\usepackage{cite}
\usepackage{siunitx}
\usepackage{upgreek}
\usepackage{amsmath, amssymb}
\usepackage[colorlinks=true, allcolors=blue]{hyperref}
\usepackage{bm}

\begin{document}

\title{A tabletop x-ray tomography instrument for nanometer-scale imaging: demonstration of the 1,000-element transition-edge sensor subarray}

\author{
    Paul~Szypryt,
    Nathan~Nakamura,
    Daniel~T.~Becker,
    Douglas~A.~Bennett, 
    Amber~L.~Dagel,
    W.~Bertrand~Doriese, 
    Joseph~W.~Fowler, 
    Johnathon~D.~Gard, 
    J.~Zachariah~Harris,
    Gene~C.~Hilton, 
    Jozsef~Imrek,
    Edward~S.~Jimenez,
    Kurt~W.~Larson,
    Zachary~H.~Levine,
    John~A.~B.~Mates, 
    D.~McArthur,
    Luis~Miaja-Avila,
    Kelsey~M.~Morgan,
    Galen~C.~O’Neil, 
    Nathan~J.~Ortiz, 
    Christine~G.~Pappas, 
    Daniel~R.~Schmidt, 
    Kyle~R.~Thompson,
    Joel~N.~Ullom,
    Leila~Vale,
    Michael~R.~Vissers,
    Christopher~Walker,
    Joel~C.~Weber,
    Abigail~L.~Wessels,
    Jason~W.~Wheeler,
    Daniel~S.~Swetz
\thanks{This research is based upon work supported by the Office of the Director of National Intelligence (ODNI), Intelligence Advanced Research Projects Activity (IARPA), through the Rapid Analysis of Various Emerging Nanoelectronics (RAVEN) research program, contract number D2021-2106170004. The views and conclusions contained herein are those of the authors and should not be interpreted as necessarily representing the official policies or endorsements, either expressed or implied, of the ODNI, IARPA, or the U.S.  Government}%
\thanks{Sandia National Laboratories is a multimission laboratory managed and operated by National Technology \& Engineering Solutions of Sandia, LLC, a wholly owned subsidiary of Honeywell International Inc., for the U.S. Department of Energy’s National Nuclear Security Administration under contract DE-NA0003525, SAND No: SAND2022-13093.}%
\thanks{P.~Szypryt, N.~Nakamura, D.~T.~Becker, J.~W.~Fowler, J.~D.~Gard, K.~M.~Morgan, N.~J.~Ortiz, C.~G.~Pappas, J.~C.~Weber, and A.~L.~Wessels are with the Department of Physics, University  of Colorado, Boulder, CO 80309 USA (e-mail: paul.szypryt@nist.gov).}
\thanks{P.~Szypryt, N.~Nakamura, D.~T.~Becker, D.~A.~Bennett, W.~B.~Doriese, J.~W.~Fowler, J.~D.~Gard, G.~C.~Hilton, J.~Imrek, Z.~H.~Levine, J.~A.~B.~Mates, L.~Miaja-Avila, K.~M.~Morgan, G.~C.~O’Neil, N.~J.~Ortiz, C.~G.~Pappas, D.~R.~Schmidt, J.~N.~Ullom, L.~Vale, M.~R..~Vissers, J.~C.~Weber, A.~L.~Wessels, and D.~S.~Swetz are with the National Institute of Standards and Technology, Boulder, CO 80305 USA.}%
\thanks{A.~L.~Dagel, J.~Zachariah~Harris, E.~S.~Jimenez, K.~W.~Larson, D.~McArthur, K.~R.~Thompson, C.~Walker, and J.~W.~Wheeler are with Sandia National Laboratories, Albuquerque, NM 87185 USA.}%
\thanks{J.~Imrek is with Theiss Research, La Jolla, CA 92037 USA.}%
}

\markboth{4EOr2C-05}{}


\maketitle

\begin{abstract}
We report on the 1,000-element transition-edge sensor (TES) x-ray spectrometer implementation of the TOMographic Circuit Analysis Tool (TOMCAT). TOMCAT combines a high spatial resolution scanning electron microscope (SEM) with a highly efficient and pixelated TES spectrometer to reconstruct three-dimensional maps of nanoscale integrated circuits (ICs). A 240-pixel prototype spectrometer was recently used to reconstruct ICs at the 130~nm technology node, but to increase imaging speed to more practical levels, the detector efficiency needs to be improved. For this reason, we are building a spectrometer that will eventually contain 3,000 TES microcalorimeters read out with microwave superconducting quantum interference device (SQUID) multiplexing, and we currently have commissioned a 1,000 TES subarray. This still represents a significant improvement from the 240-pixel system and allows us to begin characterizing the full spectrometer performance. Of the 992 maximimum available readout channels, we have yielded 818 devices, representing the largest number of TES x-ray microcalorimeters simultaneously read out to date. These microcalorimeters have been optimized for pulse speed rather than purely energy resolution, and we measure a FWHM energy resolution of 14~eV at the 8.0~keV Cu~K$\bm{\upalpha}$ line.

\end{abstract}


\section{Motivation}
As integrated circuits (ICs) become more and more complex, with state-of-the-art ICs containing feature sizes of less than 10~nm, the challenge of three-dimensionally imaging fully fabricated ICs grows appreciably. This can be important for a number of purposes, including defect analysis during process development and verification of externally fabricated components. Furthermore, due to the relative sensitivity and value of ICs that require this sort of detailed mapping, a tabletop and nondestructive solution is generally preferred in order to perform the measurement in-house and in a repeatable fashion. 

The TOMographic Circuit Analysis Tool (TOMCAT; see~\cite{nakamurasubmitted}), which originated with the Non-destructive Statistical Estimation of Nanoscale Structures and Electronics (NSENSE; see~\cite{lavely2019a, weichman2020}) concept, was developed to address these issues. Unlike previous work utilizing synchrotron ptychography to image nanoscale ICs~\cite{holler2017}, the TOMCAT approach combines a commercial scanning electron microscope (SEM) with a transition-edge sensor (TES; see~\cite{irwin1996,irwin2005}) x-ray spectrometer in a tabletop tomography instrument. Unique to the TOMCAT approach is the deposition of the x-ray generating target layer (e.g. Pt) directly on the IC under test, with a spacer layer (e.g. Si) of carefully selected thickness between the target and IC layers used to set the magnification. The electron beam of the SEM is used to generate x-rays in a small region of the target layer, which then travel through the IC and are attenuated depending on local metal density and thickness. X-rays are detected with the TES spectrometer, which has sufficient energy resolution to discern characteristic x-rays generated at the target layer from background x-rays generated elsewhere (e.g. spacer and IC layers, vacuum chamber). Although a single metal slab target has been used in all current TOMCAT measurements, a carefully designed nanopatterned target could also be used achieve x-ray generated spot sizes smaller than the electron beam spot size. Additionally, as TES microcalorimeters typically show part per thousand energy resolving power, the nanopatterned targets could also be constructed from multiple metals to \textit{multiplex} the tomography measurement. As of this writing, TOMCAT has been used to reconstruct an IC with 6 logical layers (3 wiring and 3 via) and feature sizes as small as \SI{160}{\nm}~\cite{levinesubmitted}. Voxel sizes of \SI{40}{\nm}~$\times$~\SI{40}{\nm}~$\times$~\SI{80}{\nm} were used in this reconstruction.

Because there is only so much electron current that can be packed into a sufficiently small spot size, TOMCAT's imaging speed is ultimately limited by the spectrometer's x-ray collection capacity, namely the solid angle and detector speed. The original tomography scans were done with a prototype spectrometer containing 240 TES x-ray microcalorimeters~\cite{pappas2019} read out using superconducting quantum interference device (SQUID) based time-domain multiplexing (TDM; see~\cite{doriese2016}). In order to increase the imaging speed, we are developing a new spectrometer containing 3,000 TES x-ray microcalorimeters that are about 2.5$\times$ faster than the current devices, improvements largely made possible by microwave SQUID multiplexing ($\upmu$MUX; see~\cite{mates2017,gard2018}). Completion of the full 3,000-pixel system is expected in 2023, but at this time we have built up a 1,000-pixel subarray and integrated this smaller system with the rest of the TOMCAT instrument. The goals here are twofold. First, as the mechanical integration is more or less identical in the 3,000 and 1,000 pixel implementations, much of the integration risk is retired and we can gain an understanding of expected system performance. Second, the 1,000-pixel subarray already allows us to realize a substantial imaging speed increase ($\gtrsim 4 \times$) over the original 240-pixel TDM system and can be used for tomography data collection while building the remainder of the system. This manuscript describes the 1,000-pixel implementation of the TOMCAT spectrometer, focusing on microcalorimeter characterization measurements that make useful predictions for the full 3,000-pixel spectrometer.

\section{Spectrometer Design}
\begin{figure}[ht]
	\centering
    \includegraphics[width=0.95\linewidth]{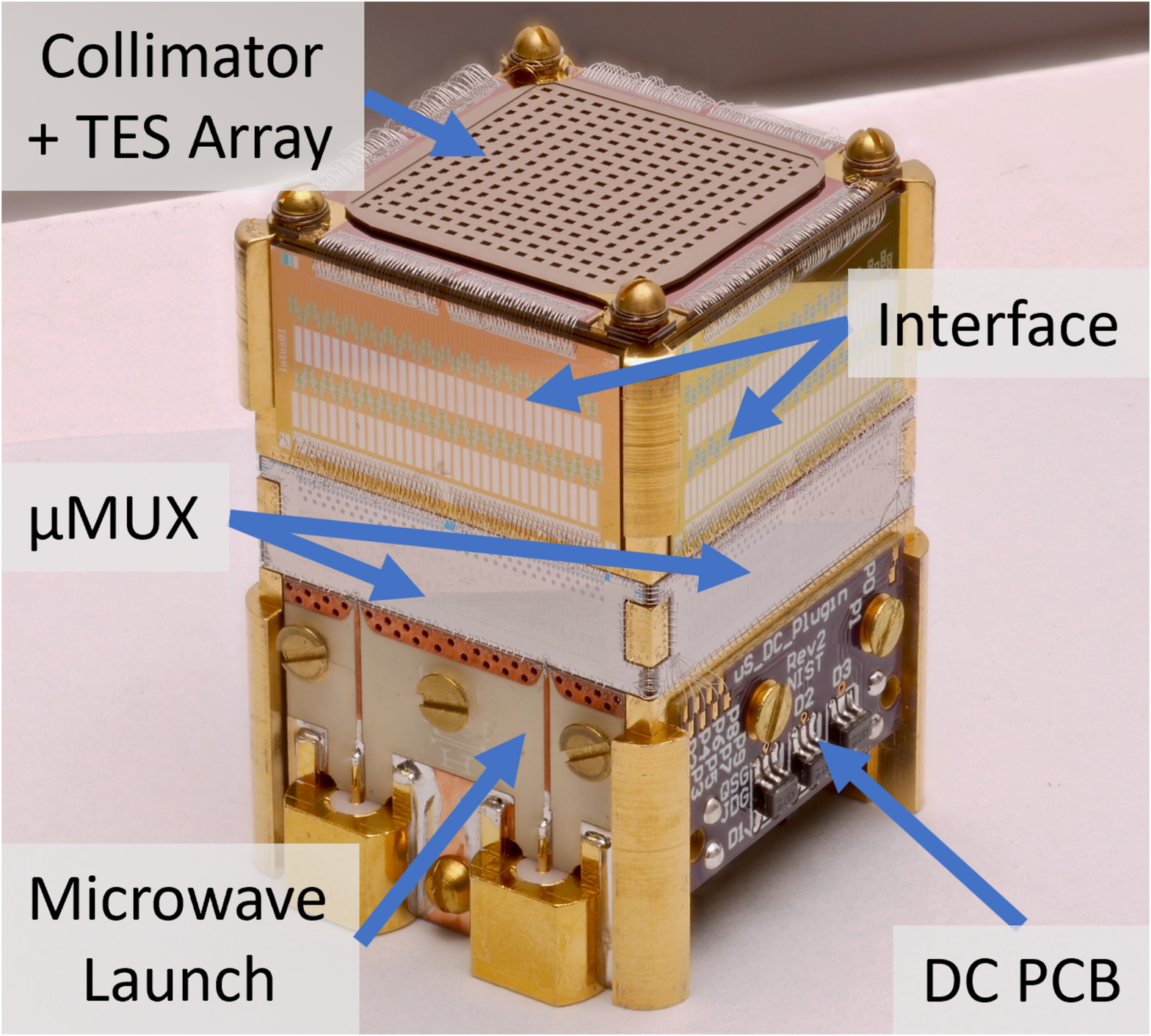}
	\caption{\label{fig:microsnout_photo}Photograph of a microsnout used in the TOMCAT spectrometer. The TES microcalorimeter array is mounted to the top of the microsnout, and a collimator (aperture) array is used to ensure x-rays are only absorbed in the microcalorimeter absorbers. A set of 4 interface chips and 4 $\upmu$MUX chips are used to bias and read out the TES microcalorimeter array. A microwave launch board and DC printed circuit board (PCB) are used to route the RF transmission line, flux ramp, and detector bias lines through the microsnout.}
\end{figure}

The 3,000-pixel (and 1,000-pixel) spectrometer differs from the prototype 240-pixel spectrometer in a few main ways. First and foremost, it uses $\upmu$MUX readout instead of TDM, providing considerably higher readout bandwidth totaling \SI{24}{GHz}. Here, there are 6 parallel readout chains each operating from \SI{4}{GHz} to \SI{8}{GHz}. The designed resonator bandwidth of each channel (pixel) is \SI{1}{MHz}, with approximately \SI{7.5}{MHz} spacing between resonators. Second, we have developed compact and modular cryogenic detector and readout packaging, referred to as the \textit{microsnout}, as shown in Fig.~\ref{fig:microsnout_photo}. Each microsnout contains a maximum of 256 detectors and associated cold readout components, and the 3,000-pixel instrument will use a total of 12 microsnouts arranged in a Swiss cross pattern. Each microsnout will use \SI{2}{GHz} of readout bandwidth, and pairs of microsnouts (\SI{4}{GHz} to \SI{6}{GHz} and \SI{6}{GHz} to \SI{8}{GHz}) will be daisy-chained together to form full readout chains. Finally, because the large detector array and cold readout electronics and wiring represent a much higher thermal load than the components in the 240-pixel spectrometer, the new spectrometer is housed in a dilution refrigerator instead of an adiabatic demagnetization refrigerator. On top of the higher cooling power of the dilution refrigerator, it has the advantage of continuous operation, further increasing data collection uptime. For a more detailed description of the 3,000-pixel spectrometer design, please see\cite{szypryt2021}.

The 1,000-pixel subarray represents a scaled down version of the 3,000-pixel system and by and large follows the same design principles. The main difference is the 1,000-pixel instrument contains a $2\times2$ array of 4 microsnouts (central 4 microsnouts of the 12 microsnout Swiss cross pattern). Additionally, each of the 4 microsnouts is designed for the \SI{4}{GHz} to \SI{6}{GHz} band, so microsnouts are not daisy-chained together in this implementation, requiring slight modifications to the room temperature readout configuration. Finally, the vacuum windows and infrared filters are also not yet in their final implementation, temporarily reducing the system efficiency.

\section{Characterization measurements}
We have begun characterizing the new TOMCAT spectrometer performance, particularly as it pertains to the 1,000-pixel implementation. We describe the performance of the cryogenic $\upmu$MUX chips in Sec.~\ref{subsec:MUX}, the 1,000-pixel microwave readout in Sec.~\ref{subsec:readout}, and the TES microcalorimeters in Sec.~\ref{subsec:TES}.

\subsection{$\mu$MUX chip performance}
\label{subsec:MUX}

A microsnout contains 4 unique $\upmu$MUX chips, each designed for a different \SI{500}{MHz} wide subband of the full \SI{4}{GHz} to \SI{6}{GHz} range. Within a $\upmu$MUX chip, there are a total of 62 readout channels (rf-SQUID and resonator), for a total of 248 usable readout channels per microsnout. This is slightly smaller than the number of TES microcalorimeters (256) to allow for some flexibility in wiring around individual problematic devices while not exceeding the resources of the microwave electronics. 

\begin{figure}[ht]
	\centering
	
	\subfloat{
        \includegraphics[width=\linewidth]{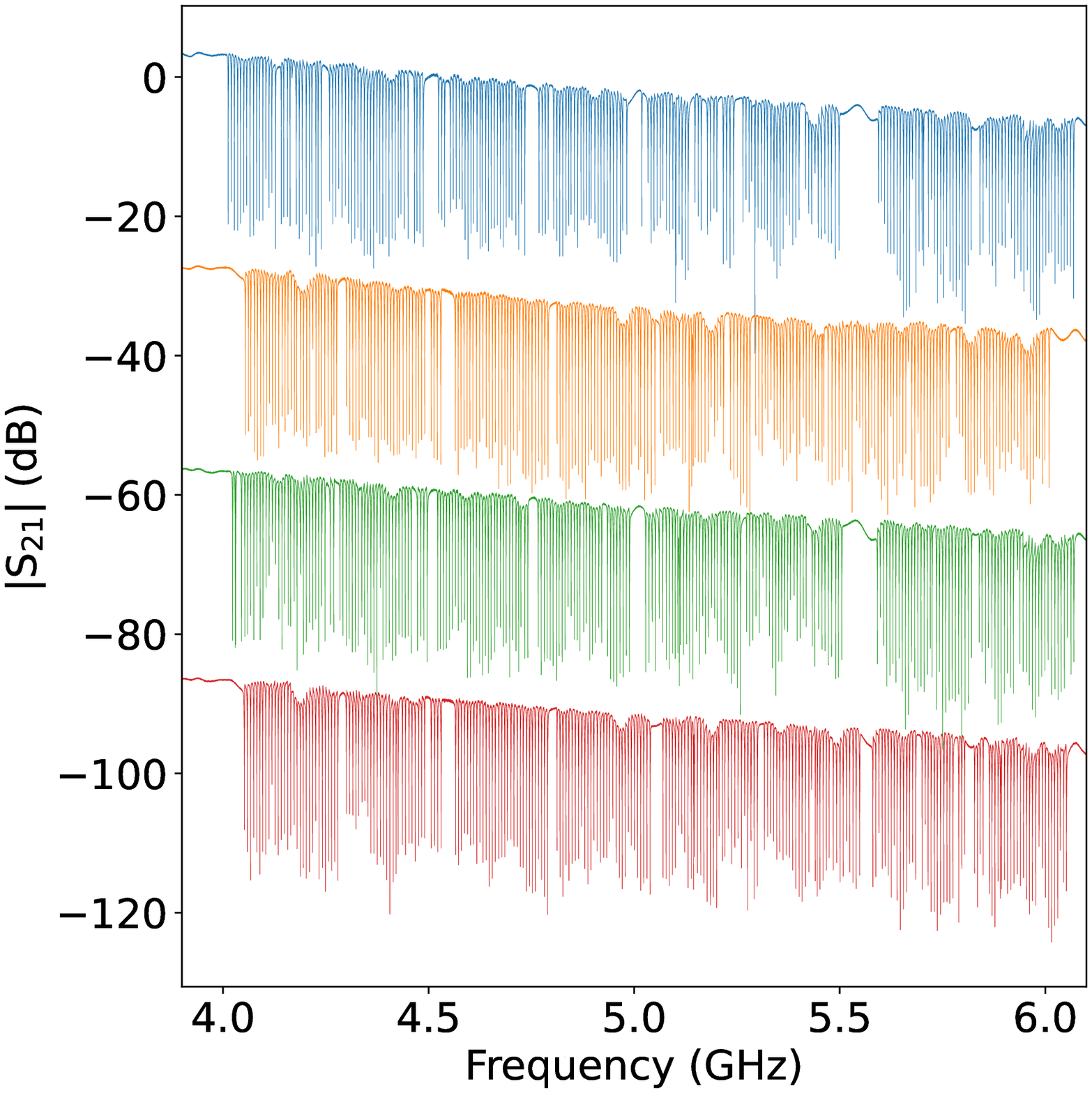}
    }
    
    \subfloat{
        \includegraphics[width=\linewidth]{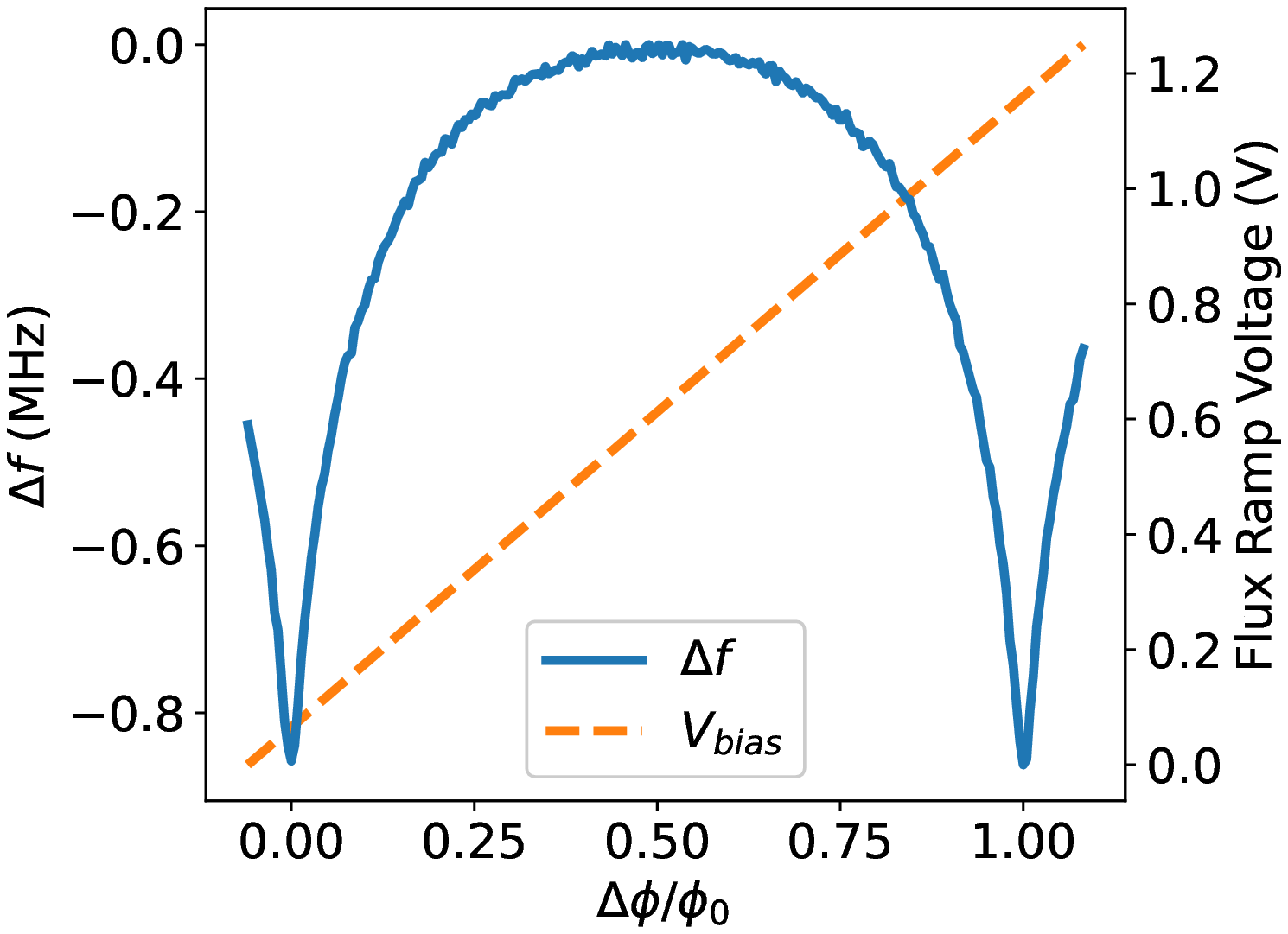}
    }
    
	\caption{\label{fig:s21}(Top) Magnitude of the complex transmission for all 4 microsnouts, as measured with a VNA. The traces of the 4 microsnouts are offset by \SI{30}{dB} steps, for clarity. A total of 967 good resonances were found in this sweep. (Bottom) Sample resonator response curve for a single readout channel when a flux ramp signal is applied to the rf-SQUID. A flux ramp amplitude of \SI{1.1}{V} corresponds to a single SQUID period.}
\end{figure}

A vector network analyzer (VNA) was used to sequentially sweep the $\upmu$MUX chips of all 4 microsnouts in the \SI{3.9}{GHz} to \SI{6.1}{GHz} frequency range, as shown see Fig.~\ref{fig:s21}~(top). Of the $4 \times 248 = 992$ potential resonances, a total of 967 good resonances (frequency roughly in expected position and adequate separation between neighboring resonators) were measured with the VNA, for a working resonator yield of \SI{97}{\%}. The VNA was also used to measure the resonator response to a flux ramp line voltage sweep on the rf-SQUIDs, as shown in Fig.~\ref{fig:s21}~(bottom). The median frequency shift caused by biasing the SQUIDs across a flux quanta was measured to be \SI{-1.1}{MHz}, comparable to the design value of \SI{-1.4}{MHz}. An additional $\sim$20 readout channels were lost due to resonators showing no or negligible modulation from the flux ramp biased SQUIDs, or for instances when this modulation caused resonator overlaps. This reduces the $\upmu$MUX chip channel yield to \SI{95}{\%}. This represents an average multiplexing factor of 237, the highest multiplexing factor achieved in TES-based x-ray microcalorimeter readout.

\subsection{4-microsnout readout demonstration}
\label{subsec:readout}

In order to read out 4 microsnouts with the microwave electronics that were already available to us, we needed to alter the readout hardware configuration. In the 3,000-pixel setup, we will use a total of 6 readout enclosures, each containing a field programmable gate array and 4 intermediate frequency (IF) boards. Each IF board covers a \SI{1}{GHz} frequency band (currently \SI{\sim950}{MHz} due to temporary firmware limitations), but there is flexibility in choosing the local oscillator (LO) frequency. A set of 4-way RF power combiners/splitters will be used to route the RF input/output signals of the 4 IF boards into a single microwave transmission line. For the 1,000-pixel implementation, we will split the resources of the readout enclosures to cover a pair of \SI{4}{GHz} to \SI{6}{GHz} bands rather than a single \SI{4}{GHz} to \SI{8}{GHz} band, which primarily involves a change to the LO frequencies and splitter/combiner setup. This way, we are able to utilize our 2 currently available enclosures for the readout of 4 microsnouts in the \SI{4}{GHz} to \SI{6}{GHz} band.

\begin{figure}[ht]
	\centering
    \includegraphics[width=0.95\linewidth]{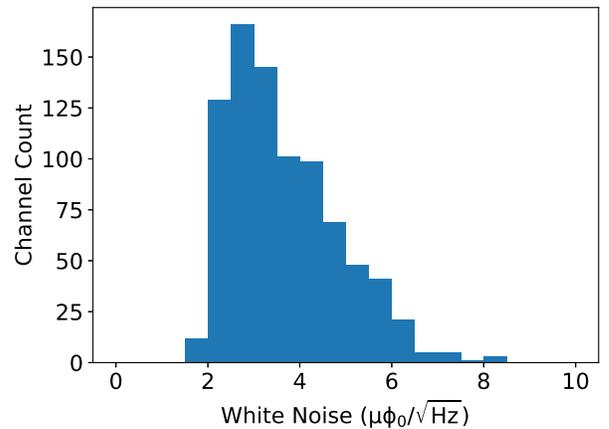}
	\caption{\label{fig:readout_noise}Histogram of readout channel white noise levels, defined as the median noise level between \SI{14.8}{kHz} to \SI{16.8}{kHz}. Detector bias was zeroed prior to collecting noise data. For each channel, we averaged 100 individual spectra with a frequency resolution of  \SI{2}{Hz}. The plotted noise is referenced to the SQUID period, and the designed coupling is \SI{10}{\upmu A / \upphi_{0}}. The median white noise level for these devices is \SI{3.4}{\upmu \upphi_{0} / \sqrt{Hz}}.}
\end{figure}

During the initial setup, the IF electronics are used to sequentially measure the the complex transmission through all 4 microsnouts and fit the center frequencies of all discovered resonators (at zero flux ramp voltage). The RF output tones are set to these frequencies. The amplitude of the flux ramp signal was set to create a $2\phi_0$ SQUID response across the flux ramp period, or \SI{2.2}{V}. The signal is sampled at 16 points during a flux ramp period, half of which are discarded to account for the flux ramp reset. A total of 951 readout channels were created during this tuning procedure, although many were incorrectly generated due to misidentified resonators and wiring shorts/opens, among other issues. Of these, 818 were found to have working detectors. Nevertheless, this is the largest number of TES x-ray microcalorimeters incorporated into a spectrometer and simultaneously read out to this date. The yield out of the 992 maximum readout channels was \SI{82}{\%}. A histogram of the readout noise in each channel is shown in Fig.~\ref{fig:readout_noise}.

\subsection{TES microcalorimeter characterization}
\label{subsec:TES}

In the TOMCAT design, the TES is a superconducting bilayer of MoAu with a target superconducting critical temperature of \SI{100}{mK}~\cite{weber2020}. The absorber is comprised of a \SI{20}{\upmu m} thick electroplated Bi layer, for high x-ray stopping power near \SI{10}{keV}, on top of \SI{2}{\upmu m} thick Au layer, for lateral thermal conductance~\cite{yan2017}. Due to the high expected count rates, the devices were optimized for fast recovery time instead of purely for energy resolution, done by increasing the thermal conductance between the TES and thermal bath.

\begin{figure}[ht]
	\centering
	
	\subfloat{
        \includegraphics[width=\linewidth]{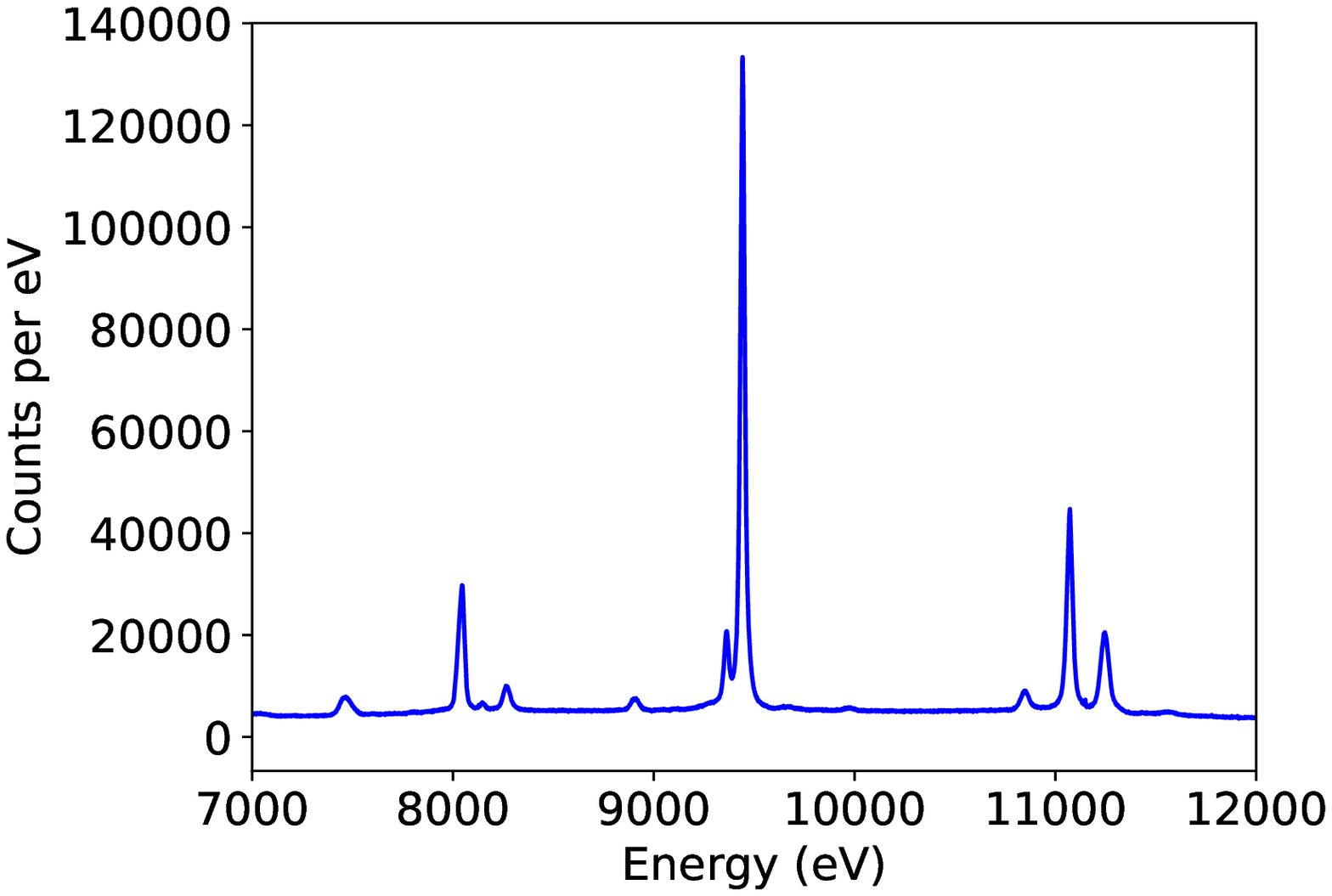}
    }
    
    \subfloat{
        \includegraphics[width=\linewidth]{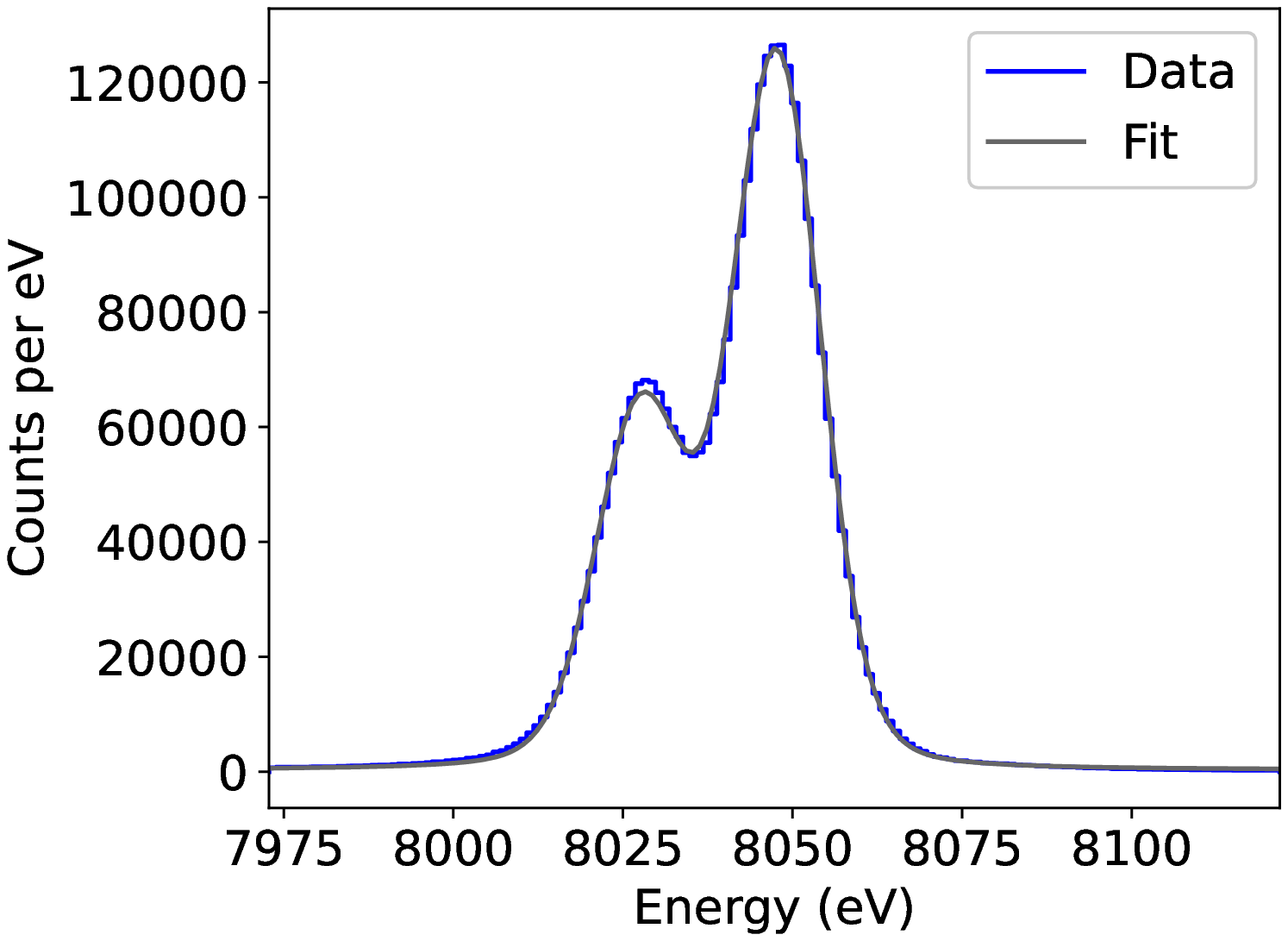}
    }
    
	\caption{\label{fig:spectrum}(Top) Coadded energy spectrum of SEM generated x-rays. The main spectral feature associated with the target layer are the Pt~L$\upalpha$ lines near \SI{9.4}{keV} and Pt~L$\upbeta$ lines near \SI{11}{keV}. These features are clearly resolvable from lines generated elsewhere in the sample (e.g. Cu~K$\upalpha$ at \SI{8.0}{keV}) or stainless steel chamber (Fe, Cr, and Ni K$\upalpha$ and K$\upbeta$ lines between \SI{5.4}{keV} to \SI{8.3}{keV}). (Bottom) Fit to the Cu~K$\upalpha$ feature in the coadded spectrum generated with external calibration source x-rays. A FWHM energy resolution of \SI{14}{eV} was extracted from the fit parameters.}
\end{figure}

During this initial set of measurements, the bath temperature was set to \SI{50}{mK} and the detectors were voltage biased to a mean level of \SI{25}{\%} of their normal state resistivity. We used the TOMCAT SEM to generate characteristic x-rays on the 4 microsnouts. X-rays were generated in the sample that was used in the tomography data collection described by \cite{nakamurasubmitted} and \cite{levinesubmitted}, with x-rays generated in the \SI{100}{nm} thick Pt target layer being of particular importance. The per pixel count rate ($\sim$\SI{40}{counts/pixel/s}) was nearly identical to that observed with the previous 240-pixel spectrometer under the same SEM conditions, expected for the current IR filter and vacuum window configuration. Because the Pt~L line structure is not known well enough to characterize detector resolution, we also used an external x-ray calibration tube source with a Cu foil target at count rates similar to that of the SEM data collection. We used the Microcalorimeter Analysis Software System (MASS; see~\cite{fowler2016, becker2019}) to convert the x-ray pulse traces to energy values. Of the initial 818 channels that were analyzed in MASS, an additional 27 were cut due to anomalous count rates or pulse shapes. A broadband energy spectrum generated with the SEM x-rays is shown in Fig.~\ref{fig:spectrum}~(top) and a fit to the Cu~K$\upalpha$ line generated with the tube source x-rays is shown in Fig.~\ref{fig:spectrum}~(bottom). A FWHM energy resolution of \SI{14}{eV} was extracted from the Cu~K$\upalpha$ fit (x-ray energy $\sim$\SI{8.0}{keV}). We note that the energy resolution predicted from the signal-to-noise ratio in the optimal filters is nearly identical in the x-rays generated by the SEM and tube source, so we do not expect degradation in achieved resolution when integrating the cryostat with the SEM.

\section{Conclusion and Future Work}
We have demonstrated the readout of a kilopixel scale TES array intended for the TOMCAT experiment. The maximum number of readout channels as set by the $\upmu$MUX hardware is 992, and of these a total of 818 channels had working devices, for a yield of \SI{82}{\%}. This is the largest number of TES x-ray microcalorimeters simultaneously read out to date. A FWHM energy resolution of \SI{14}{eV} was measured at the Cu~K$\upalpha$ line using an external calibration source. The final full-scale instrument is still in construction, but it will roughly double the potential achievable multiplexing factor (248 to 496) and triple the maximum detector count (992 to 2976). This will be an extremely powerful tool for microcircuit tomography and will undoubtedly have applications across x-ray measurement science.

\section*{Acknowledgments}
The authors would like to thank Eugene Lavely, Adam Marcinuk, Paul Moffitt, Steve O’Neill, Thomas Stark, Chris Willis and others at BAE Systems for their role in the NSENSE concept development and initial instrument integration in the NIST Boulder laboratories.

\bibliographystyle{IEEEtran}
\bibliography{bibliography_1k_tes}

\vfill

\end{document}